\documentclass{PoS}

\usepackage{bm}

\title{Skyrmions with vector mesons revisited}

\ShortTitle{Skyrmions with vector mesons revisited}

\author{\speaker{Yongseok Oh} \\
        Department of Physics, Kyungpook National University, Daegu, 702-701, Korea and \\
        Asia Pacific Center for Theoretical Physics, Pohang, Gyeongbuk 790-784, Korea \\
        E-mail: \email{yohphy@knu.ac.kr}}

\abstract{
In order to develop a model that can describe both a single baryon and multi-baryon systems on the
same footing, we re-investigate the Skyrme model in a chiral Lagrangian derived from the
hidden local symmetry (HLS) up to $O(p^4)$ including the homogeneous Wess-Zumino terms.
We use the master formulas that connect the parameters of the HLS Lagrangian and a class of
holographic QCD models, which provides a controllable way to determine the low-energy constants
of the Lagrangian once the pion decay constant and the vector meson mass are given.
Therefore, this model allows us to study the role of vector mesons in the skyrmion structure.
We find that the $\rho$ and $\omega$ vector mesons have different roles in the skyrmion structure
and that the $\omega$ meson has an important role in the properties of the nucleon.
}

\FullConference{XV International Conference on Hadron Spectroscopy-Hadron 2013\\
		4-8 November 2013\\
		Nara, Japan }

\begin{document}

\section{Introduction}

The Skyrme model is expected to provide a unified way to describe both a single baryon 
and multi-baryon systems as suggested by Refs.~\cite{BR,LPMRV03}. 
Starting with a mesonic chiral Lagrangian, the single baryon emerges as a skyrmion and 
the multi-baryon systems are simulated, for example, by putting skyrmions on a crystal lattice.

However, the previous works along this line suffer from the ambiguities in determining the
form and the low-energy constants of the effective chiral Lagrangian. 
In fact, when we consider higher-order chiral interactions or introduce more mesonic degrees of
freedom in the Lagrangian, there are too many parameters and it is not possible
to control them to understand the role of each meson in the skyrmion 
structure~\cite{FJW91,ZM94b}.

In a series of publications~\cite{MOYHLPR12,MYOH12,MHLOPR13,MHLOR13}, we have investigated
the properties of a single skyrmion and the role of vector mesons in the skyrmion structure and in
the skyrmion crystal that simulates dense baryonic matter. 
In this article, we focus on the role of vector mesons in the single skyrmion structure and 
our study on nuclear matter in this approach is reported in Ref.~\cite{MH14}.
The main idea of this approach is to start with holographic QCD models and integrate out infinite
towers of mesons except a few low-lying mesons~\cite{HMY10}. 
It then leads to a chiral Lagrangian whose low-energy constants are fixed by a few inputs through
the master formula. 
Therefore, we can control the parameters of the Lagrangian, which makes it possible to study
the skyrmion structure in a systematic way.
In this work, we use the effective Lagrangian of pion and $\rho$/$\omega$ mesons
up to $O(p^4)$ in the hidden local symmetry (HLS) scheme.
Thanks to the master formula we have only three input parameters, namely, the pion decay
constant, vector meson mass, and the HLS parameter $a$.
However, the physical quantities we are considering here are independent of $a$, so the
number of independent parameters is reduced to two.

Another point that should be addressed compared to other holographic QCD models~\cite{NHS09}
is the inclusion of the $\omega$ vector meson that can be introduced through
the homogeneous Wess-Zumino terms. In holographic QCD, this is equivalent to include 
the five-dimensional Chern-Simons term, and there is no additional free parameters.

\section{Soliton properties and the role of vector mesons}

The basic building blocks of the HLS Lagrangian are
the two $1$-forms $\hat{\alpha}_{\parallel\mu}$ and 
$\hat{\alpha}_{\perp\mu}$ defined by
\begin{eqnarray}
\hat{\alpha}_{\parallel\mu}^{} =  \frac{1}{2i} \left( D_\mu \xi_R^{}
\xi_R^\dagger + D_\mu \xi_L^{} \xi_L^\dagger \right), \qquad
\hat{\alpha}_{\perp\mu}^{}  = &\frac{1}{2i} \left( D_\mu \xi_R^{}
\xi_R^\dagger - D_\mu \xi_L^{} \xi_L^\dagger \right),
\end{eqnarray}
with the chiral fields $\xi_L^{}$ and $\xi_R^{}$, which are written in the
unitary gauge as
$
\xi_L^\dagger = \xi_R^{} \equiv \xi = e^{i\pi/2f_\pi}.
$
The vector mesons, which are the gauge bosons of the HLS, are introduced
through
\begin{equation}
D_\mu \xi_{R,L}^{} = (\partial_\mu - i V_\mu) \xi_{R,L}^{},
\end{equation}
where $V_\mu = \frac{g}{2} (\omega_\mu + \rho_\mu)$ with $g$ being the gauge coupling
constant.

Then the chiral Lagrangian up to $O(p^4)$ can be written as
\begin{eqnarray}
\mathcal{L}_{\rm HLS} & = & \mathcal{L}_{\rm (2)} +
\mathcal{L}_{\rm (4)} + \mathcal{L}_{\rm anom} ,
\label{eq:Lag_HLS}
\end{eqnarray}
where $\mathcal{L}_{(2)}$ and $\mathcal{L}_{(4)}$ are the terms of $O(p^2)$ and
$O(p^4)$, respectively, and $\mathcal{L}_{\rm anom} $ is the homogeneous Wess-Zumino
terms.
The explicit expressions for the interactions and the master formula that determines the
low-energy constants of this model 
can be found, for example, in Ref.~\cite{MYOH12}.
In this work, we use $f_\pi = 92.4$~MeV and $m_\rho^{} = 775.5$~MeV, and
we use the Sakai-Sugimoto model~\cite{SS04a}.

The soliton wave functions can be obtained by solving the equations of motion of the meson fields. 
For a single baryon that carries unit baryon number,
the solitonic solution can be found by using the following configurations,
\begin{eqnarray}
\xi(\bm{r}) = \exp\left[i\bm{\tau}\cdot\hat{\bm{r}}\frac{F(r)}{2}\right], \quad
\omega_{\mu} = W(r)\, \delta_{0\mu}, \quad \rho_0^{} = 0, \quad
\bm{\rho} = \frac{G(r)}{gr} \left( \hat{\bm{r}} \times \bm{\tau} \right)
\label{eq:hedgehog}
\end{eqnarray}
with the boundary conditions
\begin{eqnarray}
F(0) = \pi, \quad  F(\infty) = 0 , \quad
G(0) = - 2, \quad  G(\infty) = 0 , \quad
W'(0) = 0, \quad  W(\infty) = 0.
\label{eq:numprofile1}
\end{eqnarray}

In order to describe a realistic baryon of definite spin and isospin quantum numbers, 
the classical configuration should be quantized.
In this work, we follow the standard collective quantization method~\cite{ANW83}, which
transforms the meson fields as
\begin{eqnarray}
\xi(\bm{r}) &\to& \xi(\bm{r},t) = A(t)\, \xi(\bm{r}) A^\dagger(t), \nonumber\\
V_{\mu}(\bm{r}) &\to& V_{\mu}(\bm{r},t) = A(t)\, V_{\mu}(\bm{r}) A^\dagger(t),
\label{eq:mesoncollective}
\end{eqnarray}
where $A(t)$ is a time-dependent SU(2) matrix, which introduces
the angular velocity $\bm{\Omega}$ of the collective coordinate 
rotation as
\begin{eqnarray}
i \bm{\tau} \cdot \bm{\Omega} & \equiv & A^\dagger(t) \partial_0 A(t).
\end{eqnarray}
Under this rotation, the space component of the 
$\omega$ field and the time component of the $\rho$ field get excited
and their most general forms are found as~\cite{MKW87}
\begin{eqnarray}
\rho^0 (\bm{r},t) &=& A(t) \frac{2}{g}\left[ \bm{\tau} \cdot \bm{\Omega} \,
\xi_1^{}(r)
+ \hat{\bm{\tau}} \cdot \hat{\bm{r}} \, \bm{\Omega} \cdot \hat{\bm{r}} \,
\xi_2^{}(r)
\right] A^\dagger(t) , \nonumber\\
\omega^i (\bm{r},t) &=& \frac{\varphi(r)}{r} \left( \bm{\Omega} \times
\hat{\bm{r}} \right)^i ,
\label{eq:VM_excited}
\end{eqnarray}
with the boundary conditions given by
\begin{eqnarray}
\xi_1'(0)  =  \xi_1^{} (\infty) = 0 , \quad
\xi_2'(0)  =  \xi_2^{} (\infty) = 0 , \quad
\varphi(0)  =  \varphi(\infty) = 0, 
\end{eqnarray}

\begin{figure}[t]
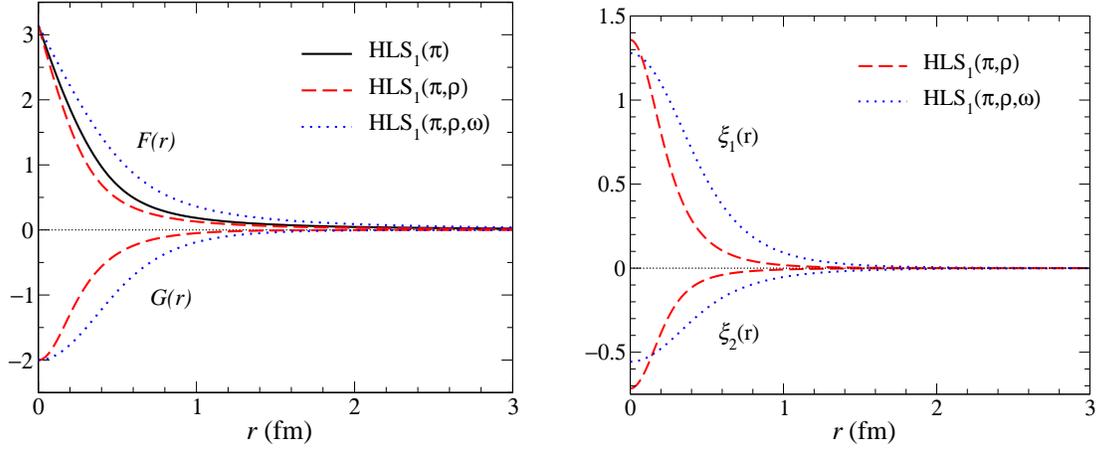
 \centering
\includegraphics[width=.45\textwidth]{fig1.eps} \qquad
\includegraphics[width=.45\textwidth]{fig2.eps} 
\caption{Comparison of the soliton wave functions $F(r)$, $G(r)$
$\xi_1^{}(r)$, and $\xi_2^{}(r)$
in the models of HLS$_1(\pi)$, HLS$_1(\pi,\rho)$, and 
HLS$_1(\pi,\rho,\omega)$, which are represented by the solid line, 
dashed lines, and dotted lines, respectively.
} 
\label{fig1}
\end{figure}

In order to understand the role of vector mesons, we consider three models.
The first is the full model in this approach that contains $\pi$, $\rho$, and $\omega$ mesons
explicitly. We call this model HLS$_1(\pi,\rho,\omega)$. 
To see the role of the $\omega$ meson, we decouple the $\omega$ meson from the full model.
This can be achieved by neglecting the homogeneous Wess-Zumino terms. 
This is the model HLS$_1(\pi,\rho)$.
Finally, to see the role of the $\rho$ meson, we consider the model HLS$_1(\pi)$ by
integrating out the $\rho$ meson in HLS$_1(\pi,\rho)$.
Then the soliton wave functions can be obtained by solving the equations of motion.
The obtained results are given in Fig.~\ref{fig1}. (The solution for the $\omega$ meson
wave function can be found in Ref.~\cite{MYOH12}.)

When the wave functions are obtained, it is straightforward to calculate the mass and radius of
a single baryon.
In Table~\ref{tab1} we give the soliton mass, the mass difference between the nucleon and the $\Delta$,
and the rms energy radius $\sqrt{\langle r^2 \rangle_E^{}}$ obtained in three models.

\begin{table}[b]
\centering
\begin{tabular}{c|ccc|cc}
\hline\hline
& HLS$_1(\pi,\rho,\omega)$ & HLS$_1(\pi,\rho)$  & HLS$_1(\pi)$  &
$O(p^2)+\omega_\mu^{} B^\mu$~\cite{MKW87} & $O(p^2)$~\cite{IJKO85} \\ \hline
$M_{\rm sol} $ & 1184 & 834 &  922 &  1407 &  1026 \\
$\Delta_M$ & 448 & 1707 & 1014 & 259 & 1131 \\ \hline
$\sqrt{\langle r^2 \rangle_E^{}}$ &  0.608 &  0.371 &  0.417  & 
0.725 &  0.422 \\
\hline\hline
\end{tabular}
\caption{Skyrmion mass and size calculated in the HLS within the Sakai-Sugimoto model with 
$a=2$.
The soliton mass $M_{\rm sol}$ and the $\Delta$-N mass difference
$\Delta_M$ are in unit of MeV while 
$\sqrt{\langle r^2 \rangle_E^{}}$ is in unit of fm.
The column of $O(p^2) + \omega_\mu^{} B^\mu$ is ``the minimal model'' of
Ref.~\cite{MKW87} and that of $O(p^2)$ corresponds to the model of
Ref.~\cite{IJKO85}.} \label{tab1}
\end{table}

From these results, we can know the followings.
\begin{itemize}
\item The inclusion of the $\rho$ meson reduces the soliton mass, which is consistent with
the claim made in Refs.~\cite{NHS09,Sutcliffe11} that the inclusion of isovector vector mesons
makes the skyrmion closer to the BPS soliton. However, we find that the inclusion of the isoscalar
$\omega$ vector meson increases the soliton mass. The different role of these mesons can be
seen in Fig.~\ref{fig1} which
shows that the $\rho$ meson shrinks the soliton wave functions, while the $\omega$ meson 
has the opposite effects.

\item In the moment of inertia, which determines the $\Delta$-$N$ mass difference $\Delta_M$ 
in the standard collective quantization, the $\rho$ and $\omega$ vector mesons have the 
opposite role again, namely, the $\rho$ meson increases $\Delta_M$, while the $\omega$ meson
decreases it. As a result, in the absence of the $\omega$ meson, $\Delta_M$ that is the
quantity of $O(1/N_c)$ becomes even larger than the soliton mass that is of $O(N_c)$, which
then causes a serious problem in the validity of the standard collective quantization method.
Therefore, the inclusion of the $\omega$ meson is important not only in phenomenology but also
to justify the standard collective quantization method.

\end{itemize}

\section{Discussion}

In summary, we have investigated the role of vector mesons in the skyrmion structure using
the HLS Lagrangian up to $O(p^4)$, which is matched by holographic QCD models by integrating 
out the vector mesons other than the lowest $\rho$ and $\omega$ vector mesons.
The parameters of the effective Lagrangian are determined by the master formula except the
pion decay constant and the vector meson mass.
In particular, we have studied the role of the $\omega$ meson in this work by including the 
anomalous parity terms. We find that the inclusion of the $\omega$ meson has an important role
not only in the properties of a single baryon but also in the justification of the use of the 
standard collective quantization method. 
The crucial role of the $\omega$ meson is also seen in the properties of baryonic matter
in the skyrmion crystal calculation as reported in Ref.~\cite{MH14}. 
However, within a chiral Lagrangian of pion only, it was shown in Ref.~\cite{MW96} that the one-loop
corrections are important to get the correct nucleon properties with the physical input parameters.
Thus the work in this direction is desired to understand the soliton structure more rigorously.

\acknowledgments 

Fruitful discussions with M.~Harada, H.~K. Lee, Y.-L.~Ma, B.-Y. Park, M.~Rho, and G.-S. Yang are
gratefully acknowledged. 
This work was supported in part by the National Research Foundation of Korea 
under Grant No.~NRF-2013R1A1A2A10007294.

\end{document}